\documentclass[aps,prd,floats,groupedaddress,twocolumn]{revtex4}
\usepackage{graphicx}

\input epsf

\def\p0{\phi_{0}}

\def\vecr{{\bf r}}
\def\veck{{\bf k}}

\long\def\comment#1{}

\def\W2{{\cal W}}

\newcommand{\bi}{B_{l_1 l_2 l_3}}

\def\ben{\begin{enumerate}}
\def\een{\end{enumerate}}
\def\bi{\begin{itemize}}
\def\ei{\end{itemize}}
\def\be{\begin{equation}}
\def\ee{\end{equation}}
\def\bea{\begin{eqnarray}}
\def\eea{\end{eqnarray}}

\def\cmm2{{\,\rm cm^{-2}}}
\def\cm2{{\,{\rm cm}^2}}
\def\cmm3{{\,{\rm cm}^{-3}}}
\def\gcmm3{{\,{\rm g\,cm^{-3}}}}

\def\fun#1#2{\lower3.6pt\vbox{\baselineskip0pt\lineskip.9pt
  \ialign{$\mathsurround=0pt#1\hfil##\hfil$\crcr#2\crcr\sim\crcr}}}

\newcommand{\ie}{{\it i.e.\,}}
\newcommand{\eg}{{\it e.g.\,}}

\begin{document}

\date{\today}
\title{Testing Gravity with the CFHTLS-Wide Cosmic Shear Survey and
  SDSS LRGs}
\author{O. Dor$\acute{e}$$^{(1)}$, M. Martig$^{(1,2)}$,
  Y. Mellier$^{(2)}$, M. Kilbinger$^{(2)}$, J. Benjamin$^{(3)}$, L. Fu$^{(2)}$, H. Hoekstra$^{(4)}$, 
   M. Schultheis$^{(6)}$, E. Semboloni$^{(2,5)}$, I. Tereno$^{(2,5)}$}
\affiliation{(1) Canadian Institute for Theoretical Astrophysics, 60 St. George St, 
University of Toronto, Toronto, ON, Canada M5S3H8\\
(2) Institut d'Astrophysique de Paris, UMR7095 CNRS,
   Universit\'e Pierre \& Marie Curie, 98 bis boulevard Arago, 75014 Paris,
   France\\
(3) University of British Columbia, 6224 Agricultural road, Vancouver,  BC, Canada V6T 1Z1\\
(4) Deparment of Physics andAstronomy, University of Victoria,
Victoria, B.C. V8P 5C2, Canada\\
(5) Argelander-Institut f\"ur Astronomie, Universit\"at Bonn, Auf dem
H\"ugel 71, 53121 Bonn, Germany\\
(6) Observatoire de Besan\c con, 41bis, avenue de l'Observatoire, BP 1615, 25010 Besan\c con Cedex, France.
}
\begin{abstract}
General relativity as one the pillar of modern cosmology has to be
thoroughly tested if we want to achieve an accurate cosmology. We
present the results from such a test on cosmological scales using cosmic
shear and galaxy clustering measurements. We parametrize potential deviation
from general relativity as a modification to the cosmological Poisson
equation. We consider two models relevant either for some linearized
theory of massive gravity or for the physics of extra-dimensions. We
use the latest observations from the CFHTLS-Wide survey and the SDSS
survey to set our constraints. We do not find any deviation from
general relativity on scales between 0.04 and 10 Mpc. We derive
constraints on the graviton mass in a restricted class of model.
\end{abstract}
\maketitle

\renewcommand{\thefootnote}{\arabic{footnote}}
\setcounter{footnote}{0}

\section{Introduction}

Observational cosmology has established the flat $\Lambda$CDM model as the
standard model of modern cosmology
\cite{Riess:2004nr,Astier:2005qq,Eisenstein:2005su,Cole:2005sx,Hinshaw:2006ia,Page:2006hz,spergel06,Tegmark:2006az}. Within
this model the energy budget of the Universe roughly goes as 
follows~: 4\% comes from baryons, 20\% from Cold Dark Matter 
(CDM) and 76\% is in the form of a cosmological constant. Those
numbers are now known to a few percent but despite those tight
constraints, the exact nature of CDM is still unknown and the high
value of the cosmological constant --or more generally Dark Energy (DE)-- keeps 
 challenging our deepest understanding of fundamental physics
 \cite{Peebles:2002gy}. The unknown nature of the required last two
 components   might cast some doubts on the foundation of this model~: the linear 
 cosmological perturbation theory in General Relativity (GR). 
 Although GR passes direct tests probing the Solar System scales ($10^{11}$m) down to
the laboratory scales ($10^{-3}$m) \cite{adelberger03}, extrapolating
its validity on more than 10 orders of magnitude to address
cosmological scales ($10^{26}$m) is questionnable. GR thus has to be
checked on all relevant cosmological scales in as many ways as
possible \cite{Peebles:2002iq}. We present in this paper such a test
on megaparsec scales using two probes of the large scale structures of
the Universe~: galaxy clustering and weak gravitational lensing.

Our approach is phenomenological and we do not aim at developing or
at testing specific fully consistent alternatives to GR. Several classes of theory have been 
studied in details in the literature with the aim at accounting for either the acceleration
of the Universe or for Dark Matter or both. The first class includes for example, a five-dimensional
alternative \cite{dvali00,Lue:2003ky} (DGP) or the addition of
non-linear terms in the Ricci scalar to the gravitational action
\cite{sawicki06,Bean:2006up,song06,Amendola:2006kh,Amendola:2006we}. The second class (for now)
includes tensor-vector-scalar theory
\cite{bekenstein04,skordis05,zhao06,chiu06,skordis06,Bourliot:2006ig,Dodelson:2006zt,Bruneton:2007si}
as a covariant theory encompassing the MOND theories
\cite{Milgrom02}. Finally, the last class includes for example the ghost condensate theory
\cite{ArkaniHamed:2003uy}. Generic models that verify Birkhoff's
theorem were also investigated in \cite{lue04}. Instead of exploring
these new theories, we rather propose simple functional forms
parametrizing the deviation from gravity on Mpc scales and constrain
them using current observational data at low redshift. If any
significant deviation from gravity was observed, we would then make
the connection with theoretical models more explicit. Our key
assumption is that the deviation from GR we explore can be written as
a modification to the Poisson equation. As such we are testing the
Poisson equation on cosmological scales.  

Several authors already tackled the tasks of constraining alternate
theories of gravity with available or planned observations
\cite{Gradwohl:1992ue,walker94,Liddle:1998ij,Choudhury:2002pu,Bernardeau:2004ar,Schimd:2004nq,knox05,linder05,Ishak:2005zs,Alam:2005pb,Sawicki:2005cc,carroll06,Knox:2006fh,Uzan:2006mf,Sereno:2006qu,stabenau06,Amendola:2007rr,bean07,Zhang:2007nk,Caldwell:2007cw,Jain:2007yk,Moffat:2007ju}. In
particular, within the context of linearized relativity, \citet{edery99} showed that it is not possible to build a 
relativistic theory that reproduces galactic phenomenology
(flat rotation curves, gravitational lensing, or dynamical mass
measurements via the virial theorem) without introducing dark
matter (their key point is that the light deflection cannot be made to
have the right sign). Following a similar approach, \citet{zhytnikov94}
consider a general metric theory that would reproduce the dynamics of
galactic systems without dark matter and reach the same
conclusion. They advocate in particular the addition of a Yukawa like potential
to the usual Newtonian one as a generic extension for a metric theory
of gravity. \citet{white01} then considered the same extension to GR
gravity, but focussed on scales relevant to weak-lensing (note that
they only discussed the effect on the deflection angle and not
on the growth rate of structure as we will do below). As they discuss,
a generic feature of such linearized extension to GR is to modify the
standard Poisson equation relating the density field to the
gravitational potential. In turn, \citet{uzan01} considered an alternative modification to the Poisson
equation that encompasses DGP theories. \citet{shirata05} and
\citet{sealfon05} constrained the Yukawa type extension to GR using
galaxy clustering measurements. Note that \citet{sealfon05} also
introduced a power-law extension to gravity. Such modifications to the
Poisson equation were later studied using numerical simulations \cite{stabenau06,bean07}.

Our work is an extension to the latter works \cite{white01,uzan01,shirata05,sealfon05,stabenau06,bean07}. In particular, we
use the latest weak gravitational lensing and clustering data to test gravity on Mpc
scales at low redshift. The two alternate theories of
gravity we constrain are  introduced in Sec.~\ref{sec:theoretical_motivation}, 
before detailing their phenomenology. We then describe our data-sets
and methodology in Sec.~\ref{sec:methodo} before presenting and
discussing our results in Sec.~\ref{sec:results}.

\section{Modified theories of gravity and constraint methods}
\label{sec:theoretical_motivation}

\subsection{Theoretical motivations}

To allow for deviations from general relativity opens up many
possibilities and a wide range of gravity theories that we cannot explore 
 exhaustively. We restrict ourselves to a few somewhat phenomenological models that offer
both a physical motivation and observationaly tractable cosmological
implications. 

We will consider two different models that we   
 introduce below: the Yukawa  and the Uzan-Bernardeau type models.

As \citet{white01}, we follow \citet{zhytnikov94} who within well defined
postulates presents some general arguments regarding the shape of a
linearized metric theory of gravitation.  These postulates comprise post-Newtonian slow motion extension ($(v/c)^2\ll 1$) and weak gravitational field regime 
 relevant to the scale we consider in this work. Their description
 includes forces mediated by massless or massive scalar and tensor modes. The metric reduces to
\bea
g_{00}^{} & = & (-1+2\Phi) \\
g_{ij}^{} & = & (+1+2\Phi)\delta_{ij}
\eea
where the potential $\Phi$ is given by 
\bea
\Phi(\vecr) & = & (1-\alpha)\Phi(\vecr,0)+\alpha \Phi(\vecr,m) \\
\Phi(\vecr,m)   & = & G\int {\rho(\vecr')d^3\vecr'\over
  |\vecr-\vecr'|}e_{}^{-m|\vecr-\vecr'|}.\label{eq:pot_yuk}
\eea
Thus, to the usual Newtonian potential $\Phi(\vecr,0)$ a Yukawa type
potential $\Phi(\vecr,m)$ is added corresponding to propagating massive modes.

Considering the evolution of over-density, $\delta(\vecr,t)$ or its
Fourier transform, $\tilde\delta(\veck,t)$ the standard linear theory
of perturbations leads to the Poisson equation relating the
gravitational potential to the over-density. The Poisson equation
writes in comoving coordinates and Fourier space \cite{peebles80}
\be
\tilde\Phi(\veck,a) = -{3\over 2} {H_0^2\Omega_{m0}\over
  a}{\tilde\delta(\veck,t)\over k^2} f(\veck,a) 
\ee
where $a$ is the scale factor, $H_0$ is the Hubble constant,
$\Omega_{m0}$ the present matter density. $f$ is a deterministic
function equal to 1 for the standard gravity. It can easily be
shown that the inclusion of a Yukawa type potential as in
Eq.~\ref{eq:pot_yuk} leads to 
\be 
f^{Yuk}(\veck) \equiv f(\veck) = 1 - \alpha{1 \over 1+\left({k\over a \ m}\right)^2}.
\label{eq:f_yuk}
\ee
As such, exploring alternative theories of gravity will be equivalent
for us to test the Poisson equation.

Observational constraints on similar models (but with slightly
different notations) coming from galaxy surveys \cite{shirata05,sealfon05}, 
 as well as using also their non-linear evolution using N-body simulations \cite{stabenau06,bean07},  have already been
studied. Note that our
notation matches that of \citet{white01} and corresponds to that of
\citet{shirata05} and \citet{sealfon05} with ($\alpha\rightarrow -\alpha$,
$\lambda\rightarrow-1/m$) and \cite{stabenau06} with ($\alpha\rightarrow -\alpha$, $\lambda\rightarrow-1/r_s$).

Another set of models captures, in the context of superstring theories,
some brane induced phenomenology. In such a scenario, a generic feature 
seems to be the existence of two scales below/above which standard gravity is altered
\cite{Randall:1999vf,Binetruy:1999ut,Kogan:2000xc,Gregory:2000jc}. The
smaller scale (of order a millimeter or less), irrelevant to our measurements, corresponds to the
existence of Kaluza-Klein gravitons. On the other hand, above the
branes separation scale, we also expect gravity to be altered. Since
this scale is exponentially larger than the previous one, it becomes
cosmological. The use of large scale structures as a probe of this form
of deviation from standard gravity as been advocated by \cite{binetruy01,uzan01,lue04}, who include the model of
\cite{Gregory:2000jc}. \citet{uzan01} focus in particular on
weak gravitational lensing,  as such we will follow their work more
closely. They describe in real space the violation of Newton's law above
a given physical scale, $r_s$, as a multiplicative function to the standard Newtonian
potential, $\Phi(r,0)$,
\bea
\Phi(r) = \Phi(r,0){1\over 1 +{r\over r_s}}
\eea
so that on small scales, $r\ll r_s$, we recover the usual Newtonian
gravity. As before, this translates into a modification to the Poisson
equation with a multiplicative function, $f^{UB}$ \cite{uzan01,bean07}
\bea
\lefteqn{f^{UB}(k,a,r_s) \equiv f(k,a) }&\\
& = {kr_s\over
  2a}\left[-2\sin({kr_s/a})\displaystyle{ \int_{kr_s/a}^{\infty}{\cos(t)\over t}dt} \right.\nonumber\\
&+ \left.\cos(kr_s/a)\left(\pi-2\displaystyle{\int_0^{kr_s/a}{\sin(t)\over t}dt} \right)\right]
\label{eq:f_ub}.
\eea
$r_s$ is an arbitrary scale  and is the parameter we will
constrain later on. 

\begin{figure*}[t!]
\begin{center}
\includegraphics[angle=90,width=0.49\textwidth]{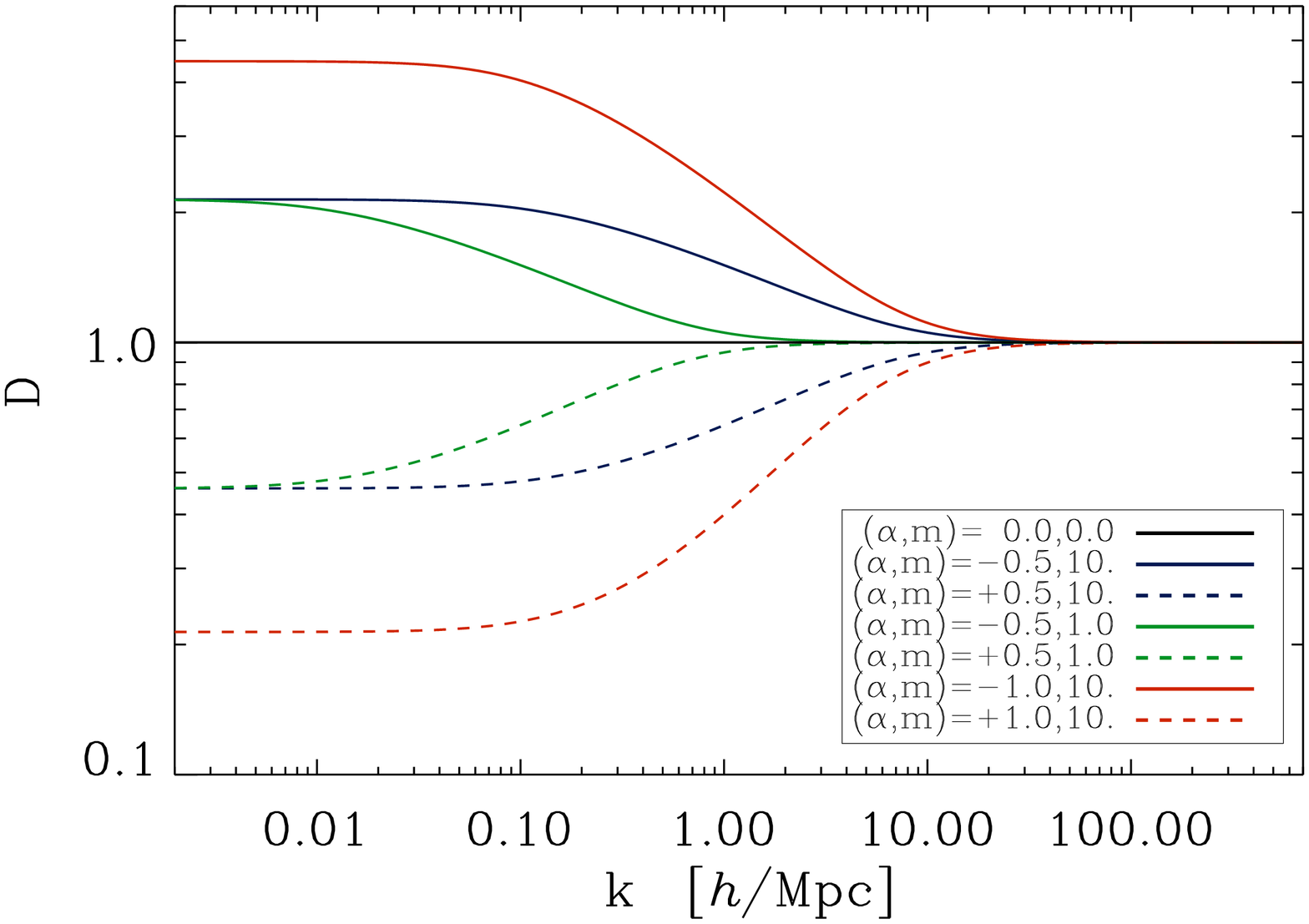}
\includegraphics[angle=90,width=0.49\textwidth]{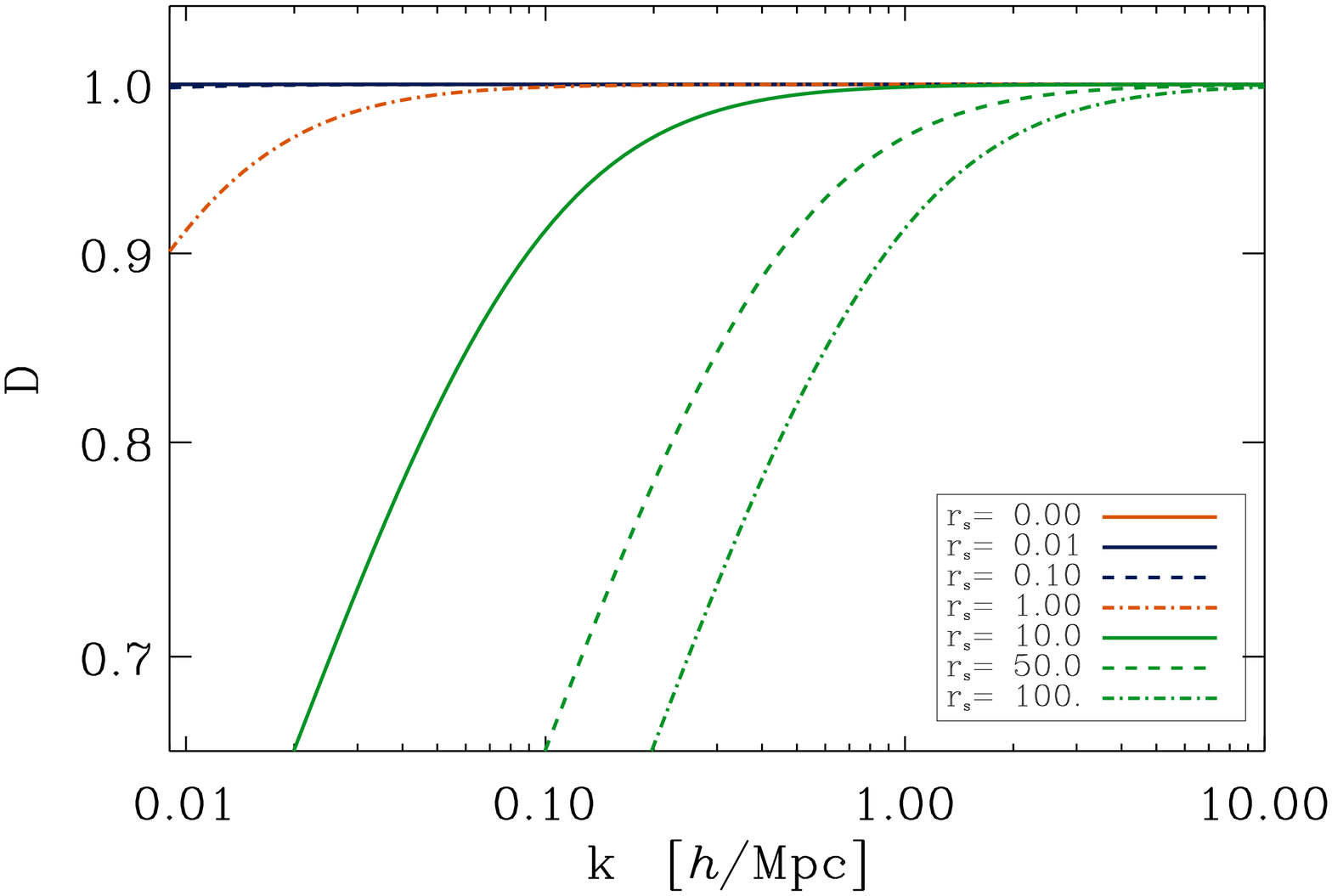}
\end{center}
\caption{Ratio of linear growth factor  of modified gravity versus
  $\Lambda$CDM at $z=0$. From left to right, the Yukawa model and the UB model.}
\label{fig:ratio_D}
\end{figure*}

From now on, we will focus on these two models  (Yukawa and UB)
and study their phenomenological implications on cosmological scales
($\simeq$ 10 Mpc) before constraining their free parameters. 
We deliberately restrict our conclusions from those models to the
scales we are probing, \ie we do not explore the parameter space where
$1/m$ or $r_s$ are close to the Horizon scale. Note that we choose $1/m$
and $r_s$ to be physical distances, \ie we do not allow them to be
comoving distances.


\subsection{Phenomenology}

Modifications to the Poisson equation as described above translate into an
alteration of gravity above a characteristic scale ($1/m$ or $r_s$). To
illustrate this we choose to focus on the linear perturbation theory and
compute the linear growth rate of structure. We do not attempt at
building a fully covariant theory of gravity (a task well beyond the
scope of this paper) and therefore we consider that the evolution of
the background metric will be the one of the $\Lambda$CDM model
currently favored by the data (See however \cite{dvali03,carroll05}
for a discussion of background evolutions). Our motivation to do so
relies on the fact that any alternative theory of gravity will have to
reproduce the now well observationally established $\Lambda$CDM
background evolution. It seems thus fair to keep using it as an
effective description of the evolution of the background
metric. However, if our test of gravity on 10 Mpc scales or so were to
detect any deviation, a more acute theoretical description would
be required. 

A modification to the Poisson equation
leads to a change on the linear growth rate of perturbations.
 If $D$ is defined so that $\delta(\veck,t) = D(\veck,t)\delta(\veck,t_0)$
where $t_0$ is an arbitrary original time \cite{peebles80,shirata05}, then it satisfies the equation
\bea
\ddot D + 2H\dot D = {3\over 2}{H_0^2\Omega_{m0}\over a^3}f(k)D
\label{eq:d_diff}
\eea
where\ " $\dot{}$\ " \  denotes time derivative and where we restrict ourselves
to time independent $f$ and $k$ is the comoving wave number. The Hubble
 parameter, $H$, and its evolution are defined as for the $\Lambda$CDM
standard case by 
\bea
H & = & {\dot a \over a} \\
H^2(a)  & = & H^2_0\left({\Omega_{m0}^{}\over a^3} +
\Omega_\Lambda^{}\right)\; .
\eea
Solving  Eq.~\ref{eq:d_diff} numerically is straightforward. To fix 
boundary conditions, by analogy with the standard case, we consider
the limit $a\rightarrow 0$ and look for a solution $D=a^n$. Using the
fact that $H^2\rightarrow H_0^2\Omega_0/a^3$ (we tend to Einstein-de
Sitter) when $a\rightarrow 0$ we find that the growing mode solution
is given by 
\bea
n & = & {1\over 4} \left(\sqrt{1+24f(k)} -1\right)\mbox{ if }
1+24f(k)\ge 0.
\eea

Numerical solutions to this equation at $z=0$ for various parameters are
plotted in Fig.~\ref{fig:ratio_D}, or more precisely the growth  rate for
modified gravity divided by the expected growth rate of structures for
a $\Lambda$CDM model. Let's discuss the Yukawa model first (left panel). As
visible in Eq.~\ref{eq:pot_yuk}, the Yukawa potential introduces a
damping on gravity for $r\gg 1/m$ which entails a slow down of
structure growth. More generally, an increase or a decrease of gravity is expected, 
 depending on the sign of $\alpha$. This is
visible in the left panel of Fig.~\ref{fig:ratio_D} for various $\alpha$s and $m$s. By
construction, our models converge toward standard gravity at the 
smallest scales ($f\rightarrow 1$ when $k\rightarrow\infty$) and tends
to a weaker but scale independent gravity for ${k\over ma}\ll 1$ hence
the flat limit for low $k$.  The amplitude of this plateau depends 
only on $\alpha$. The on-set of the transition is however controlled
by both $\alpha$ and $m$ and, for a given $\alpha$, the smaller $m$
the lower in $k$ the transition occurs.   For a given $m$ however, the
smaller $|\alpha|$, the lower in $k$ is the departure from standard
gravity. Those results agree with previous published results, as \eg
\citet{stabenau06} and the analytical estimates of \citet{Sereno:2006qu}.

The second model we consider is the UB model, as defined in
Eq.~\ref{eq:f_ub}. It only depends on the cut-off scale $r_s$ and as
expected, the growth of structures for scales larger than $r_s$ will be
slower as compared to $\Lambda$CDM as the gravity gets weaker. This
is clearly visible in the second panel of Fig.~\ref{fig:ratio_D}.

\section{Analysis methodology}
\label{sec:methodo}

\subsection{Observables}

Making use of the previously derived linear theory predictions, we now
constrain our models using two probes of the growth rate of large scale
structures: cosmic shear and galaxy clustering. 

For the former we will use the latest Canada-France-Hawaii-Telescope
Legacy Survey (CFHTLS \footnote{{\texttt
    http://www.cfht.hawaii.edu/Science/CFHLS}}) observations as in
\citet{fu07}, a work extending the  analyses of the previous release
\cite{sembolini06,hoekstra06,Benjamin:2007ys}. For the latter we will
use the Sloan Digital Sky Survey (SDSS \footnote{{\texttt
    http://www.sdss.org}}) matter power spectrum estimated from the
clustering of Luminous Red Galaxies as in \citet{tegmark04}.  

Since weak gravitational lensing provides a mean to directly image
the total mass distribution as a function of redshift, it is potentially a
 powerful way to constrain accurately the growth of structures and thus any
theory affecting it. The data we use here are based on the
recently published analysis of the CFHTLS-Wide survey (release T0003)  
 that spreads over 57
square degrees (34.2 after masking). The depth of the weak lensing 
  catalog reaches a magnitude of
$i'_{AB}=24.5$ and corresponds to a  galaxy density of about 13.3
gal./arcmin$^2$.  It comprises an effective  
sample of about 1.7$\times 10^6$ galaxies whose shape correlation properties
have been analysed by \citet{fu07}. 
 Although five bands will
eventually be available, only the $i'$ band was used for this analysis. 
The area used to produce this data-set is about twice larger 
than the previous release of the CFHTLS (but only 35$\%$ of the total size of the survey) 
  but explores much larger scales. The width of this
area turns out to be crucial to our analysis since the cosmic shear is now 
measured from $1$ arcmin up to $4$ degree, well into the linear
regime. Since we are looking for variation in the shape of the power
spectrum, this wide range of scales is particularly valuable. Although it
would be highly beneficial to our project, no tomography measurements
has been carried out with these data yet. Systematic effects are
constrained to be smaller than statistical errors. ``B-modes'' in
particular are negligible at all scales we use. To constrain gravity, we choose two lensing statistics whose
properties are slightly different~: the shear E/B correlation
functions, $\xi_E$, and the compensated filter known as  aperture mass statistic,
$M_{ap}$ \cite{Schneider:2001af,fu07}. The former is defined as 
\bea
\xi_E(\theta)  =  {1\over 2\pi}\int_0^\infty dk\ kP_{\kappa}^{}(k)J_0(k\theta)
\label{eq:xi_def}
\eea 
whereas the latter is defined as 
\bea
M_{ap}^2(\theta)  =  {288\over \pi\theta^4}\int_0^\infty {dk\over
  k^3}\ P_{\kappa}^{}(k)J_4^2(k\theta)
\label{eq:map_def}
\eea
where
\bea
P_{\kappa}(\ell) &=& {9H_0^4\Omega^2\over4c^4}\int_0^{\omega_H}d\omega
\left({W(\omega)\over a(\omega)} \right)^2
D(k,w)P_{\delta}\left(\ell/f_K(\ell)\right).\nonumber\\
&&
\label{eq:pkappa}
\eea
$P_{\delta}$ is the density power spectrum at $z=0$ and the geometry factor $W$ is defined as 
\bea
W(\omega) = \int_w^{w_H} dw'n(w'){f_K(w'-w)\over f_K(w')}
\eea
with the redshift distribution  given by
\bea
n(z) = {\beta\over z_s\Gamma\left({1+\alpha\over
    \beta}\right)}\left({z\over z_s}\right)^\alpha \exp\left[-\left({z\over z_s}\right)^\beta\right]
\label{eq:nz}
\eea
and where $f_K$ is the comoving angular diameter distance. Both statistics involve a different weighting of the
convergence power spectrum depending on whether a wide or narrow kernel
is favored in real or Fourier space \cite{VanWaerbeke:2003uq}. As
illustrated in \cite{fu07}, the use of two different statistics
provide an extra consistency check for the measurements and their
interpretation. 

We assumed here that the estimators have been properly calibrated and
that $\xi_+=\xi_E$ ( in the notations of
\cite{fu07,Benjamin:2007ys}). As \citet{fu07} we will discard the four
smallest angular scales when using the $M_{ap}^2$ statistic due to
excessive E/B  mixing. We will use the exponential redshift distribution defined
in Eq.~\ref{eq:nz} as \citet{Benjamin:2007ys}.
 It contains three free parameters
$z_s$, $\alpha$ and $\beta$ that are determined using the photometric redshift calibrated on the VIRMOS VLT
Deep Survey \cite{Ilbert:2006dp}. The uncertainties in $\alpha$ and $\beta$ are sufficiently small that we can fix $\alpha$
and $\beta$ to respectively 0.838 and 3.43. However we still need to marginalize
over the $z_s$ uncertainties,  as it will be described in the next sub-section.  
  The choice of the  \citet{Benjamin:2007ys} redshift distribution instead of the \citet{fu07} seems 
  somewhat inconsistent and may look like a useless complication.  It is primarily motivated by the huge computational gain in keeping 
 $\alpha$ and $\beta$ constant. Although the  \citet{Benjamin:2007ys}
  distribution is sligthly less accurate than the power-law one
  used in \citet{fu07}, the few percent difference is not relevant for
  our purpose but it would be if we were trying \eg to constrain the
  overall amplitude instead of marginalizing over it.

The large-scale real-space power spectrum $P_\delta(k)$ is measured using a
sample of ~2$\times 10^6$ galaxies from the Sloan Digital Sky Survey, covering
about 2$\times 10^3$ square degrees with median redshift $z\simeq 0.35$. We follow the
methodology described in \citet{tegmark04} and use the likelihood
code available at \footnote{\texttt{http://www.hep.upenn.edu/~max/sdsspower.html}}. Note
that these measurements of the real-space matter power spectrum $P(k)$
are up to an unknown overall multiplicative bias factor, $b$, over which we
will marginalize as described below.  This bias is assumed to be
  linear, \ie constant, in the range probed by the SDSS and CFHTLS-Wide data sets. A strong scale
  dependance, although unlikely on those scales, could potentially be degenerate
  with the effects we are looking at.

One subtlety arises when computing the density power spectrum required
to  derive theoretical expectations for these observables. In
both cases, to define an initial power spectrum we use the fitting
formula of \citet{novosyadlyj99} as used by \citet{tegmark04}. We
normalize it using $A_s$, the amplitude of density fluctuations at
$k=0.05$ Mpc$^{-1}$ (see \cite{Spergel:2003cb}). We then modify the growth
rate as a function of $k$ and $a$ according to the numerical results
of Eq.~\ref{eq:d_diff}. Since the galaxy clustering data we are using
are in the mildly non-linear regime ($k\le 0.2h/$Mpc), we do not apply any non-linearity corrections besides
the redshift space distortion one as in \citet{tegmark04}. 
The situation is different for cosmic shear. Schematically, the
smallest angular scale of interest to our lensing data is
1. arcmin. Neglecting projection effects, this angular scale
corresponds to about 0.6 Mpc/$h$ at the median redshift of our 
lens population ($z=0.5$), that is around $k\simeq 2h/$Mpc. At these
scales, the non-linear corrections to the power spectrum are of order
a few and thus non-negligible. However, in the standard $\Lambda$CDM model, 
 these corrections are computed using fitting formulae calibrated on
numerical simulations \cite{peacock96,smith03} that  use standard gravity.
 Their relevance to the alternative theory of gravity we consider is  
 therefore not obvious.  This loop hole has already been tackled in
the literature and does not turn out to be a critical issue. \citet{stabenau06} used numerical simulations
with modified gravity and found that in the context of our Yukawa model, both 
 prescriptions give reasonable fits. \citet{peacock96} seems to provide a better fit for
negative $\alpha$ (our conventions), whereas for positive $\alpha$,
both prescriptions are as good. These results have recently been confirmed and
extended to include the \citet{uzan01} model we consider
\cite{bean07}. As a precautionary measure, all our results will use
both prescriptions. 

\begin{figure*}[!t]
\begin{center}
\includegraphics[height=0.49\textwidth,angle=90]{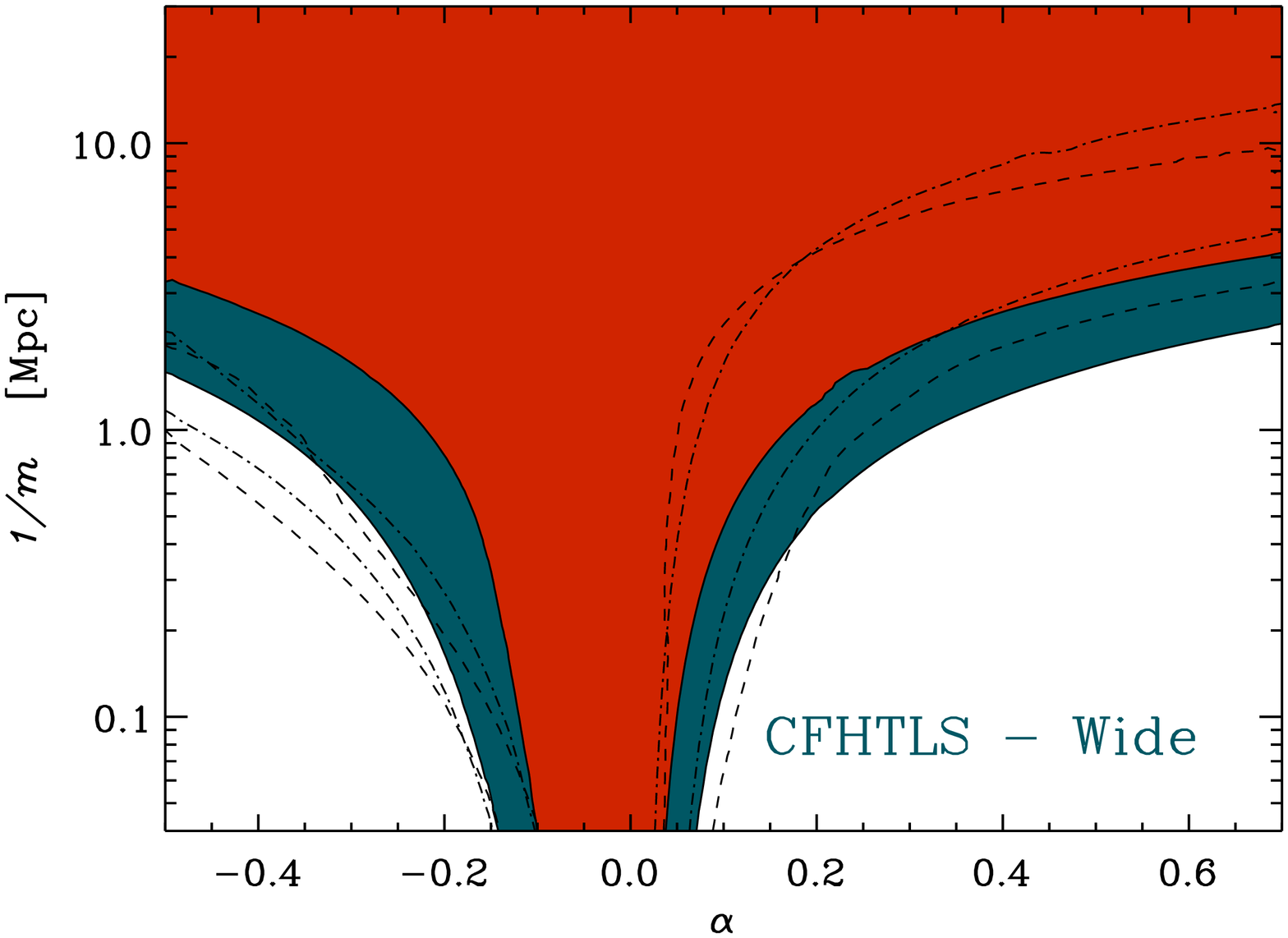}
\includegraphics[height=0.49\textwidth,angle=90]{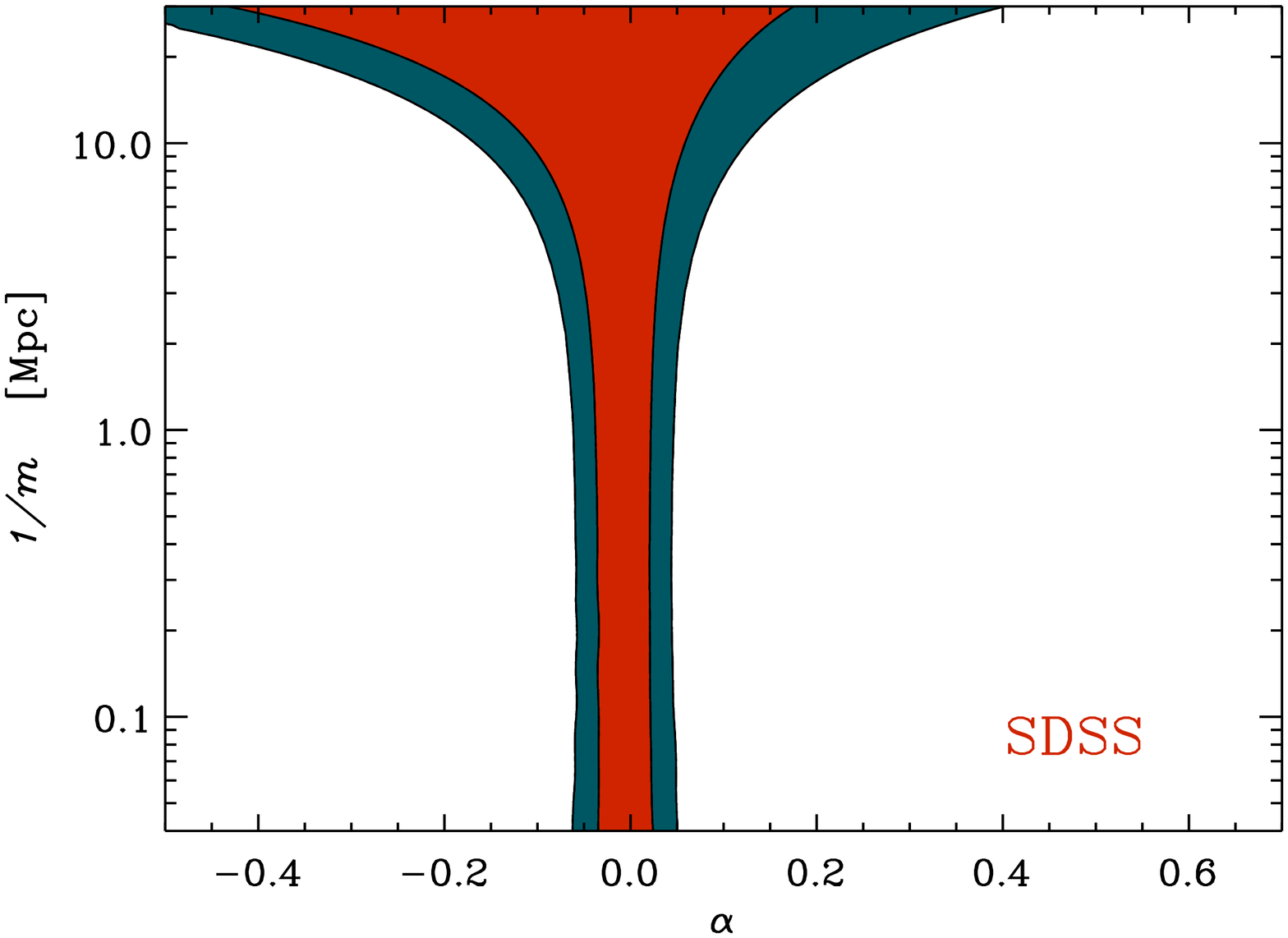}
\end{center}
\caption{Likelihood contours at 68\% and 95\% confidence levels for the
  $1/m$ and $\alpha$ parameters of the Yukawa type modification to
  gravity. The left panel corresponds to the CFHTLS-Wide constrains
  while the right panel corresponds to SDSS LRGs. Colored contours
  correspond for CFTHSL-Wide to the use of the $M_{ap}^2$ statistic
  with the {\texttt halofit} non-linear prescription. The dashed lines
  were obtained using the $M_{ap}^2$ statistic with the
  \citet{peacock96} prescription whereas the dot-dashed lines were obtained with
  the {\texttt halofit} non-linear prescription but using the $\xi_E$
  statistic. The agreement between these various prescriptions and
  statistics is a satisfying of robustness of our measurement. As
  expected given the wider area covered by SDSS (42 times bigger than
  the current status of CFHTLS-Wide), the SDSS constraints are much
  narrower despite the bias uncertainty.}
\label{fig:contours_wl_lrg}
\end{figure*}

\subsection{Likelihood methodology and priors}
\label{sec:like_prior}

We set up to constrain these alternative theories of gravity using maximum
likelihood methods. We choose to maximize the likelihood in a five (four)
dimensional parameter space. Two (one) of these parameters are the prime physical parameters we are interested in, be it
either $(\alpha,m)$ in the Yukawa case ($(r_s)$ in the UB
case). We then choose an amplitude parameter $A_s$ corresponding to the
normalization of density perturbations at $k$=0.05Mpc$^{-1}$. This
parameter is tightly constrained by WMAP 3 year measurements only, but
using WMAP as a prior seems inappropriate in our case. The CMB
constraints come from much larger scales than the upper limits of CFHTLS-Wide and SDSS data, 
 typically a few hundreds of Mpc/$h$. Since our modified gravity models affect gravity and so
the matter power spectrum on scales larger than a characteristic
scale -- be it $1/m$ or $r_s$-- that far exceed the 0.1-10 Mpc range of the data,
it is not suitable to apply a CMB prior on $A_s$. Instead, it is  
 preferable to normalize the power spectrum on smaller scales. Such a
normalization is available through the joint measurement of the $Ly_\alpha$
forest flux fluctuations with other constraints on the linear matter
density power spectrum \eg SDSS \cite{McDonald:2004xn}. 
 \citet{Seljak:2006bg} showed that at the 2$\sigma$ level, the WMAP 3
 years and small scale $Ly_\alpha$ constraints are consistent. As
 such, we will use conservatively a prior on $A_s$ derived from the WMAP and $Ly_\alpha$ 
 joint measurements~: we assume a uniform prior for $A_s$
in between the 99.7\%CL of WMAP 3 year only \cite{spergel06} : $
2.988\le$ $\log(10^{10}A_s(k=0.002\rm{\, Mpc}^{-1}))$$\le3.324$. 
 Note that because of the altered growth of structures, the scaling
$\sigma_8$ with $A_s$ will depend on  $(\alpha,m)$ or $(r_s)$. 
This choice of normalization does not matter for the  SDSS data, because of the unknown bias, but is more critical for weak-lensing
observations. 

Finally, we have to take into account two extra nuisance
parameters. For weak lensing observables, we marginalize over the
parameter $z_s$ of the source distribution (Eq.~\ref{eq:nz}) assuming a
uniform prior within the 2$\sigma$ constraints of \citet{Benjamin:2007ys}~:
$z_s=1.172\pm0.026$. We also marginalize the SDSS likelihood over the
bias parameter assuming a uniform prior $0.25<b<4$ which comprises most reasonable solutions \cite{shirata05}. 

The limited range in scales of our observations leads us to restrict
ourselves to the following range for our prime parameters. We explore
$-0.5<\alpha<0.7$ and $0.001<m<30$ in the case of Yukawa, and $0.001<r_s<10$ in the case of UB. 

Finally, we fix the other cosmological parameters to their best fit
flat $\Lambda$CDM WMAP value only (\citet{spergel06}). Namely  we consider
$h=0.732$, $\omega_m=0.1277$, $\omega_b=0.0223$ and $n_s=0.958$. We
did not explore the dependance of our results with regards to those
parameters although we expect it to be small given the small
uncertainties with which those parameters are now measured
\cite{sealfon05}. Those parameters essentially fix the
background evolution and the initial power spectrum.

We explore the likelihood in this five dimensional space using a
regular linear gridding of this hyper-volume (except for the $m$
dimension that we explore using a logarithmic binning), with typically ~100 samples per
direction. Our code makes use of the existence of fast dimensions $b$ that require very little computation to be explored so
that we can explore this full hyper-volume in 10 hours or so using 8 1.6GHz Intel Xeon cpus.

\section{Results and discussions}
\label{sec:results}

Fig.~\ref{fig:contours_wl_lrg} displays the 2D 68\% and 95\%
contour levels for the Yukawa model $m^{-1}$ and $\alpha$ parameters
using CFHTLS-Wide (left panel) and SDSS LRGs (right panel). Other
parameters have been marginalized over. Focusing on the CFHTLS panel
first, we display several set of contours. The colored contours
correspond to  the $M_{ap}^2$ statistic (Eq.~\ref{eq:map_def}) with the \citet{smith03} \texttt{halofit}
prescription to model non-linearities. The long dashed lines correspond 
to the $M_{ap}^2$ statistic with the \citet{peacock96} prescription. The dot-dashed lines correspond 
to the $\xi_E$ statistic (Eq.~\ref{eq:xi_def}) with the  
\citet{smith03} \texttt{halofit} prescription. To the level of
accuracy we are interested in this work, this figure shows that those various
methods and measurements are consistent with one another. This is a
reassuring statement regarding the robustness of our constraints. We
will from now on quote numbers from the $M_{ap}^2$ statistic using the
\texttt{halofit} prescription (colored contours).

\begin{figure}[t!]
\begin{center}
\includegraphics[height=0.49\textwidth,angle=90]{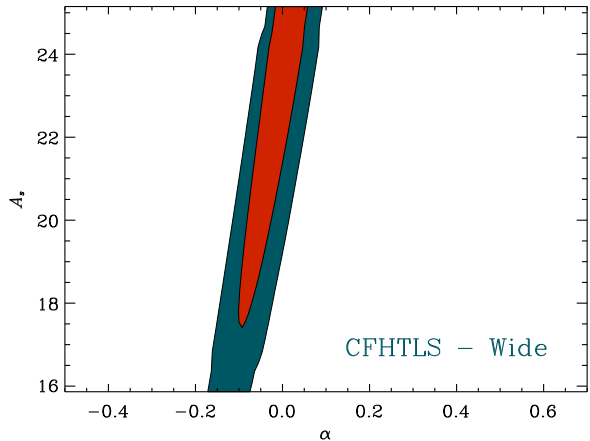}
\end{center}
\caption{Likelihood contours at 68\% and 95\% confidence levels for the
  normalisation parameter $A_s$ and $\alpha$ parameter of the Yukawa type modification to
  gravity. We consider for this plot only $1/m \le 0.1$Mpc so that we
  explore the lower part of the left plot in
  Fig.~\ref{fig:contours_wl_lrg}. This plot illustrate the degeneracy
  between $\alpha$ and $A_s$ that is expected given
  Fig.~\ref{fig:ratio_D} where we can see that for a given $m$ and redshift,
  $\alpha$ will change the amplitude of $P(k)$ above a given scale. }
\label{fig:2d_alpha_As}
\end{figure}

In the case of Yukawa type models, for a given $1/m$ the deviation from gravity
increases with increasing $|\alpha |$ (see Fig.~\ref{fig:ratio_D})
hence it is expected constraints be centered on our favored  value 
 $\alpha=0$.  Therefore, Fig.~\ref{fig:contours_wl_lrg} 
   first shows that no deviation from standard gravity is favored by CFHTLS
data. Since weak-lensing constrains small scales best (small $1/m$),
we observe narrower constraints at lower $1/m$.  
The broadening of the
contours at high $1/m$ shows the limits of our data in terms of large
scale sensitivity. Below this broadening, say below $1/m \le 0.5$ Mpc,
we can see that $\alpha$ and $1/m$ are decorrelated, \ie the
constraints on $\alpha$ seem independent of $1/m$ and vice
versa. This can be understood as follows. For a given $1/m$, the
transition to modified gravity is set and $\alpha$ will modify the
amplitude of the effect on the linear growth rate as visible in the
right panel of Fig.~\ref{fig:ratio_D}. Since for weak-lensing the relevant
quantity is a weighted projection of the linearly evolved power
spectrum (with non-linear corrections applied for each $z$ considered), we expect a degeneracy
between $\alpha$ and the overall amplitude of the power spectrum set by
$A_s$. This is illustrated in Fig.~\ref{fig:2d_alpha_As} where we plot
the 2D contours in the $\alpha-A_s$ plane for $1/m \le 0.1$Mpc. The
range of $A_s$ shown here corresponds to our uniform prior on $A_s$
motivated in section \ref{sec:like_prior}. Clearly this choice of
prior is a key for weak-lensing measurement and tightening it more 
would also strengthen our constraints on $\alpha$  but we choose a fairly
conservative prior in order to secure the consistency with the latest
CMB and $Ly_\alpha$ observations.
 
The right panel of Fig.~\ref{fig:contours_wl_lrg} shows the same
constraints using the SDSS LRGs. Clearly, they are much
tighter. The reason for this difference can be understood in the 
following way. Since we marginalize over the amplitude of the power
spectrum within our $A_s$ priors (and also on the bias in the case of
SDSS), our test of gravity basically relies on constraining the shape of the power
spectrum. Because the SDSS data we are using cover about 2400 square
degrees with accurate photo-$z$ for all of those (and a median redshift
of $z\simeq 0.1$), the power spectrum we are using covers the range
0.02 $h$/Mpc$\le k \le$0.3$h$/Mpc. On the other hand, CFHTLS-Wide
data, makes uses of only 57 square degrees so that the errors on the
same mode are at least 6 times larger due to cosmic variance only. In
addition, since  cosmic shear is sensitive to a weighted
projection of the matter power spectrum (see Eq.~\ref{eq:pkappa}), the shape information is
partially erased due to projection effects. This effect could be
greatly alleviated if we were able to separate the lenses into various $z$
planes, \ie to perform tomography, but such a work is still in progress for the CFHTLS survey.

\begin{figure}[!t]
\begin{center}
\includegraphics[height=0.5\textwidth,angle=90]{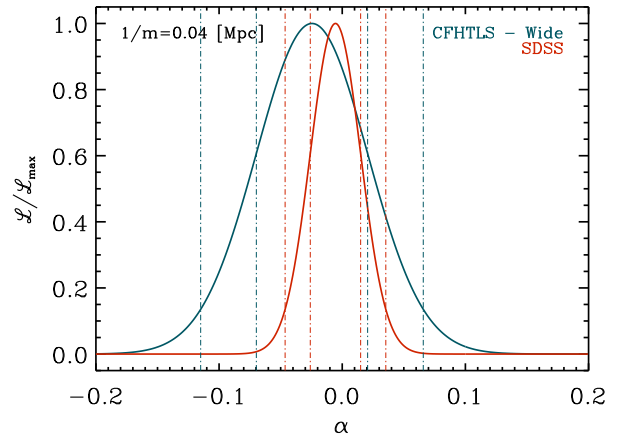}
\end{center}
\caption{1D likelihood distribution for $\alpha$ when one sets
  $1/m=0.04$Mpc. Likelihood corresponds to either CFHTLS-Wide
  or SDSS. Vertical lines correspond to 68\% and 95\% confidence
  levels. We can see that for this scale, CFHTLS-Wide constraints are
  only $\simeq$2 times worse than SDSS ones. This can be understood
  from the fact that lensing is sensitive to the overall normalization
  of the power spectrum.}
\label{fig:1d_alpha}
\end{figure}

On the other hand, for a given scale $m$ well probed by both cosmic
shear and galaxy clustering data, say $m=$25 Mpc$^{-1}$ ($1/m=$0.04
Mpc) CFHTLS performs better than one would expect from simple
cosmic variance arguments as is visible in
Fig.~\ref{fig:1d_alpha}. This figure shows the 1D marginalized
likelihood distribution for $\alpha$ when one sets $1/m=$0.04
Mpc. It can be understood as a horizontal slice in
Fig.~\ref{fig:contours_wl_lrg} for $1/m=$0.04 Mpc. We see in
this figure that the 68\% uncertainty level for $\alpha$ using
CFHTLS-Wide is only 2.2 bigger than the SDSS ones. This comes from the
fact that lensing does not suffer from the bias uncertainty so that
within our amplitude prior, lensing is sensitive to the overall
amplitude whereas galaxy clustering is not. On the other hand, as
illustrated above, if we broaden our normalization prior then the lensing constraints will
broaden too.

\begin{figure}[!t]
\begin{center}
\includegraphics[height=0.5\textwidth,angle=90]{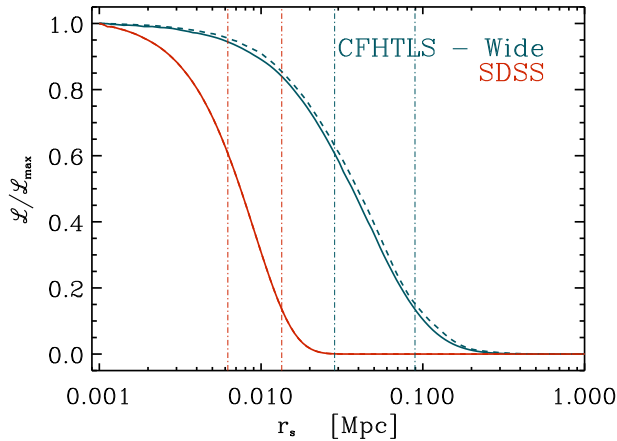}
\end{center}
\caption{$r_s$ likelihood distribution for the UB model. Vertical
  lines correspond to 68\% and 95\% confidence levels. Blue lines
  correspond to CFHLS-Wide constraints whereas red lines correspond to
  SDSS. The blue dashed line corresponds to the use of the
  \citet{peacock96} prescription to model non-linearities while the
  solid line corresponds to the \citet{smith03} \texttt{halofit} prescription.}
\label{fig:contours_uab_wl_lrg}
\end{figure}

Fig.~\ref{fig:contours_uab_wl_lrg} displays the 1D 68\% and 95\%
confidence levels on $r_s$ from the UB model obtained either with the cosmic shear or
galaxy clustering observations. Once again, we do not see any evidence
for any deviation from standard gravity on the scales we probed. Not surprisingly, we find that as for
the Yukawa model, SDSS performs better than CFHTLS in constraining $r_s$ 
by about one order of magnitude. Once again, this stems directly from
  the wider range of scales probed by SDSS.

Our results for the Yukawa model are consistent with those previously
published in the literature using the SDSS results
\cite{shirata05,sealfon05} or a simpler form of cosmic shear
measurements \cite{white01}. Our constraints on the UB model are in
agreement with those derived using the shear 3-point function \cite{Bernardeau:2004ar}. 

Before discussing further our results, we reassess the theoretical hypothesis underlying our work. First we
  assume that the background evolution is identical to the
  $\Lambda$CDM one, which might over-estimate the effects on large
  scales \cite{uzan01}. Second, we assume that the relation between
  the deflection angle and the gravitational potential that
  underlies the convergence power spectrum definition in
  Eq.~\ref{eq:pkappa} is unchanged as compared to GR. Models were this
  relation does not hold are as such neglected in our approach
  \cite{Schimd:2004nq,Uzan:2006mf,song06,Amendola:2007rr}. Third, we
  consider deviation from the usual cosmological Poisson equation only
  and restricted ourselves to some parametric form. Those limitations
  come from the fact that we choose not to work within the frame of a
  well defined theory of gravity. As such, our approach may be
  described as a test of the Poisson equation in a $\Lambda$CDM
  cosmology.

We did not study explicitly the dependence of our constraints as a
function of other parameters. \citet{sealfon05} showed that
their variations within currently observationally allowed bonds had minor effects in
this context. Besides, the current constraints are
such that most of the major degeneracies of the $\Lambda$CDM are
broken. The likelihoods are thus almost Gaussian  (see Fig. 10 of
\citet{spergel06}) and marginalizing over a parameter is very close to
setting it at its most likely value. In contrast, however,
 the effects of massive neutrino would be degenerated with 
 gravity ones we explore since they also modify the slope of the power spectrum at high
$k$ \cite{Sereno:2006qu}.  On the other hand, as explained before, we also did not quote
any constraints in terms of $\sigma_8$ since this number is obviously
very model dependent in our case and a comparison with the standard
$\Lambda$CDM value would not be meaningful. 

The main conclusion of our work is that we did not find any sign of
deviation from gravity using large scales probes of the matter
perturbations on cosmological scales (between 0.04 and 10 Mpc) at low
redshift. Whereas probes of gravity on those scales at low redshifts
are scarce, there is another important one we did not use, namely the
Integrated Sachs-Wolfe (ISW) effect \cite{Sachs:1967er}. However, since our
modifications of gravity are totally ad-hoc and do not correspond to
any covariant theory, it is unclear yet how meaningful would be our
extension to the scales relevant for ISW and we leave that for
future work.

Another indirect implication of our work is the fact that the dark
matter clustering evolution as probed by cosmic shear measurements,
\ie the deflection of lights and the clustering of luminous matter
   as probed by galaxy surveys provide
consistent constraints on theories of modified gravity on 0.04-10 Mpc
scales. This agreement has some more quantitative applications that we
leave for future work. Note also that within our models, our
constraints imply that galactic scale physics is unchanged. Whereas
the Yukawa model we discussed was originally introduced as an
alternative to dark matter on galactic scales, the extrapolation of our constraint
suggests that galactic physics is still ruled by GR.

It is worth noticing that in the context of the linearized theory of
gravity, the addition of a Yukawa term corresponds to setting a non-zero mass to
the graviton \cite{misner73,Kogan:2000xc,Choudhury:2002pu}. Our
current results do not allow us to place constraint on $1/m$ if we leave
$\alpha$ vary freely. If we were to set $\alpha =1$ as in
\citet{Choudhury:2002pu}, then we obtain the following constraints on
the mass of the graviton $1/m \ge 5.71$Mpc (95\% CL) using CFHTLS, and
$1/m \ge 67.6$Mpc using SDSS;  that is $m \le 1.11\times 10^{-30}$eV using
CFHTLS and $m \le 9.40\times 10^{-32}$eV using SDSS.   

Further improvements on tests of gravity on cosmological
  scales should explore more physical assumptions as well as estimators, like tomography. 
  SDSS data show the sky coverage is of primary importance, while CFHTLS-Wide data 
 reveal this survey still covers too narrow a part of the sky and should go one order of magnitude in terms of angular scales. 
  This will be possible soon, with the advent of next generation large scale cosmic shear surveys,
  such as \eg DES, DUNE, LSST or SNAP \footnote{{\texttt http://www.lsst.org}\\{\texttt
      http://cosmology.uiuc.edu/DES/}\\{\texttt
      http://www.dune-mission.net/}\\{\texttt http://snap.lbl.gov/}}.

\acknowledgements{}
We are grateful to Rachel Bean, Francis Bernardeau, Neal Dalal,
Catherine Heymans, Mike Kesden, Justin Khoury, Itsvan Laszlo, Patrick
McDonald, Scott Tremaine and Licia Verde for useful discussions and
insightful remarks, and to Jean-Philippe Uzan for a careful reading
of this paper. LF thanks the ``European Association for Research in Astronomy" training site (EARA) 
 and the European  Commission Programme for the Marie Curie Doctoral Fellowship  
 MEST-CT-2004-504604. LF, MK and YM thank the CNRS-Institut National des Sciences 
 de l'Univers (INSU) and  the French Programme National de Cosmologie (PNC) for 
 their support to the  CFHTLS cosmic shear program. IT and YM acknowledge the support 
 of the European Commission Programme 6$^{th}$ framework, Marie Curie
  Training and Research Network ``DUEL'', contract number
  MRTN-CT-2006-036133. IT thanks the Deutsche Forschungsgemeinschaft under  
 the project SCHN 342/8--1 and the Priority Programme 1177.
  MK is supported by the CNRS ANR ``ECOSSTAT'',
  contract number ANR-05-BLAN-0283-04.  HH is supported
  by the Natural Sciences and Engineering Research Council (NSERC),
  the Canadian Institute for Advanced Research (CIAR) and the Canadian
  Foundation for Innovation (CFI). 
 This work is based in part  on observations obtained with MegaPrime/MegaCam, a joint 
 project of CFHT and CEA/DAPNIA, at the Canada-France-Hawaii Telescope (CFHT) which 
is operated by the National Research Council (NRC) of Canada, the Institut National 
 des Science de l'Univers of the Centre National de la Recherche Scientifique (CNRS) 
 of France, and the University of Hawaii. This work is based in part on data products 
 produced at TERAPIX and the Canadian Astronomy Data Centre as part of the 
 Canada-France-Hawaii Telescope Legacy Survey, a collaborative project of NRC and CNRS.

\bibliography{modgrav}

\end{document}